\documentclass{aa}
\usepackage{graphicx}
\usepackage{epsfig}
\usepackage{bm}

\usepackage[varg]{txfonts}
\usepackage{url}
\usepackage{hyperref}
\usepackage{color}
\usepackage{orcidlink}
\usepackage{color}
\usepackage{amsmath}

\newcommand{\ixpe}{\text{IXPE}\xspace}
\newcommand{\extp}{\text{eXTP}\xspace}

\newcommand{\unit}[1]{\mbox{\boldmath $\hat{#1}$}}

\newcommand{\beq}{\begin{eqnarray}}
\newcommand{\eeq}{\end{eqnarray}}
\newcommand{\be}{\begin{equation}}
\newcommand{\ee}{\end{equation}}

\newcommand{\rmd}{{\rm d}}

\begin{document}

\title{Polarized radiation from the spreading layer \\ of weakly magnetized neutron stars} 
\authorrunning{A. Bobrikova et al.}
\titlerunning{Polarization from the spreading layer of the WMNS}

\author{Anna~Bobrikova\inst{1}\orcidlink{0009-0009-3183-9742}
\and  Juri~Poutanen\inst{1}\orcidlink{0000-0002-0983-0049}
\and  Vladislav~Loktev\inst{1,2}\orcidlink{0000-0001-6894-871X}
}

\institute{Department of Physics and Astronomy, 20014 University of Turku, Finland \\ 
\email{anna.a.bobrikova@utu.fi}
\and Department of Physics, P.O. Box 64, 00014 University of Helsinki, Finland\\
}

\date{Received 24 September 2024 / Accepted 6 March 2025}

\abstract{Observations show that the X-ray emission of the accreting weakly magnetized neutron stars is polarized.  
We developed a theoretical model for polarized radiation from the spreading layer, which is the extension of the accretion flow boundary layer onto the neutron star surface. 
We calculated the Stokes parameters of the radiation and accounted for relativistic aberration and gravitational light bending in the Schwarzschild metric.
We show that regardless of the geometry, the polarization degree of the spreading layer does not exceed 1.5\%.  
Our results have implications with regard to the understanding of the X-ray polarization from weakly magnetized neutron stars observed with the {Imaging X-ray Polarimetry Explorer} and the future {enhanced X-ray Timing and Polarimetry} mission. }


\maketitle
%

\section{Introduction}

Weakly magnetized neutron stars (WMNSs) in low-mass X-ray binary systems (LMXBs) are among the brightest X-ray sources. 
The neutron stars (NSs) in these systems emit mostly by accretion of matter that leaks from the companion star that fills its Roche lobe. 
The emission of these sources is highly variable due to the outbursts and rapid changes in the geometry of the system and is a subject of thorough research.

The emission mechanism of WMNSs has been studied for decades. Spectroscopic results obtained in X-ray and radio bands allowed scientists to conclude that the emission mostly comes from the accretion disk that formed around the NS and from a Comptonized component, which can be associated with a boundary layer (BL) between the disk and the NS surface \citep{Shakura88} or a spreading layer (SL), which is a layer of accreted matter at the NS surface \citep[see, e.g.,][]{Lapidus1985,IS99}. 
Timing observations added more information about the QPOs in these systems. Observable both at Hz and kHz frequencies  \citep{vdKlis1989,vdK00}, they are currently associated with the events that occur near the NS surface. 
However, the exact geometry of the sources remains unknown. 

Polarimetric studies add two more measurable variables to the data sets we obtain from the sources. With the launch of the Imaging X-ray Polarimetry Explorer \citep[\ixpe;][]{Weisskopf2022} in December 2021, we are capable for the first time of measuring the polarization degree (PD) and polarization angle (PA) of the WMNS X-ray emission with high precision. 
These new observables can shed new light on the emission mechanisms, the geometry of the source, and the interactions of the light on the way from the emitting region to the observer. 

Fifteen WMNSs have been observed with \ixpe in the three years of operation \citep[see][for a recent review]{Ursini24}, each providing answers and new questions. From an upper limit in GS~1826$-$238 \citep{Capitanio23} to a $10\sigma$ polarization detection in GX~340+0 \citep{LaMonaca2024b,Bhargava24}, we saw a broad variety of complex phenomena in all the sources. For some of the sources, we were able to compare the recent observations with previous results: in Cyg~X-2 \citep{Farinelli23}, the agreement between \ixpe and Orbiting Solar Observatory~8 \citep[OSO-8;][]{Weisskopf76,Long80} was reported, while in Sco~X-1 \citep{LaMonaca2024}, a discrepancy between the archive and recent observations was observed. 
The PD increased with energy in GX~9+9 \citep{Ursini2023}, 4U~1624$-$49 \citep{Saade24}, and 4U~1820$-$303 \citep{DiMarco23}, and the PD increases from the soft to the hard state in XTE 1701$-$462 \citep{Cocchi23} and GX~5$-$1 \citep{Fabiani24}. 
Significant variability of the PA with time and/or energy was observed in Cir~X-1 \citep{Rankin2024} and GX~13+1 \citep{Bobrikova24,Bobrikova24b,DiMarco2025}. 
Various interpretations, such as scattering in the disk wind, tilted rotating NS, and precession of the rotation axis, were suggested for all these phenomena. Further investigation is still required, however. 

To understand the polarimetric data, we need to have expectations regarding the PA and PD values of the emission coming from different parts of the source, such as the accretion disk, the BL, and the SL. 
The polarization of the disk emission was studied in detail \citep[see, e.g.,][]{2008MNRAS.391...32D, 2009ApJ...691..847L, LVP22}. 
\citet{Lapidus1985} estimated the SL emission and its polarization. 

In the current article, we present a detailed study of the polarized emission coming from the SL. 
We calculate the emission of the rapidly rotating SL accounting for relativistic aberration and light bending in the  Schwarzschild metric. 
We account for rotation of the polarization plane due to relativistic effects  \citep{poutanen20}. 
Unlike \citet{Lapidus1985}, we account for the rapid motion of the matter in the SL and consider various geometries of the SL. 

The more recent studies that were inspired by \ixpe observations, such as the study by \citet{2022MNRAS.514.2561G}, applied relativistic Monte Carlo codes to  simulate the polarimetric properties of the emission regions of various geometries. 
We present a semi-analytical approach to the problem, similar to that of \citet{Farinelli24}. 
However, we study a broader range of parameters, use a more realistic approach to the properties of the layer such as velocity and temperature, and investigate the dependence of the polarization on energy. 

The remainder of the paper is organized as follows. 
In Sect.~\ref{sec:theory}, we present the model that describes the formation of polarized radiation in a SL at the NS surface. 
In Sect.~\ref{sec:Results}, we apply the formalism to different geometrical configurations of the SL and evaluate the polarimetric properties of the emitted radiation. 
In Sect.~\ref{sec:Results_em}, we study the effect of varying the emission model. 
We compare our results to the existing observations of the WMNS by \ixpe in Sect.~\ref{sec:IXPE}. 
We conclude in Sect.~\ref{sec:Summary}. 

\section{Emission model} 
\label{sec:theory} 

\begin{figure}
\begin{center}
\includegraphics[width=0.5\textwidth]{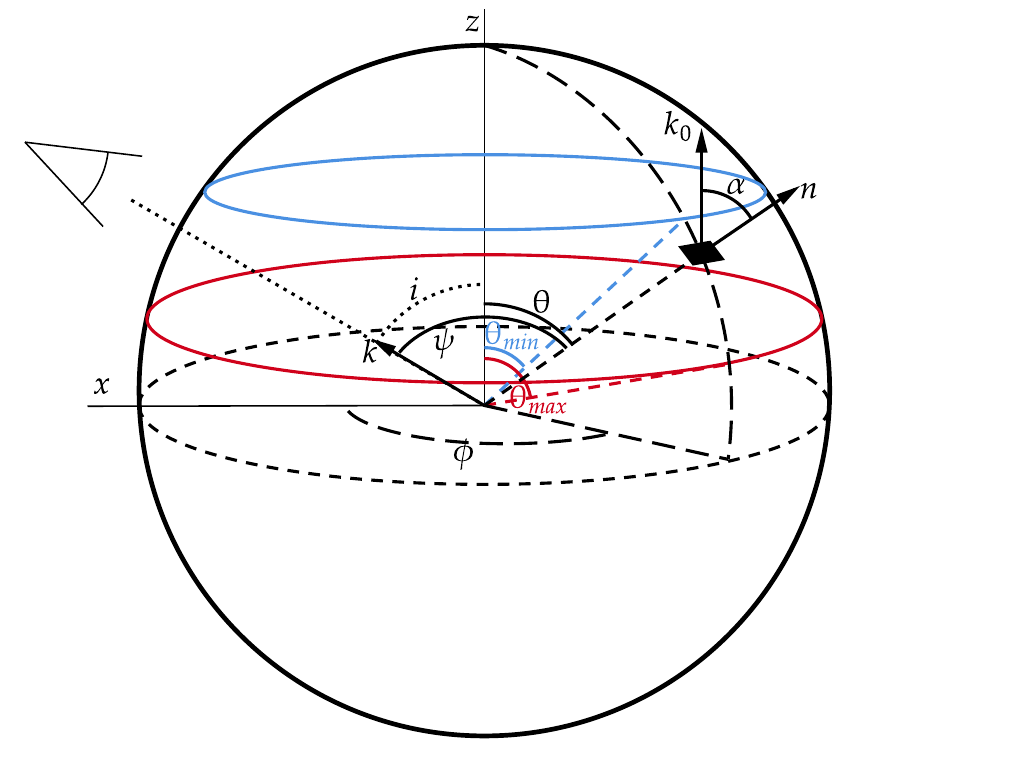}
\caption{Geometry of the problem. The SL is confined between the two circles with the colatitude $\theta_{\rm min}$ and $\theta_{\rm max}$, plotted in blue and red, respectively.}
\label{fig:geom}
\end{center}
\end{figure}

We considered a spherical NS of radius, $R$, and mass, $M$, with a SL covering part of the NS surface. We defined the layer as a belt located between the minimum and maximum colatitude, $\theta_{\rm min}$ and $\theta_{\rm max}$ (see Fig.~\ref{fig:geom}). The coordinate system was chosen such that the z-axis was aligned with the rotation axis of the NS, and the disk lay in the xy-plane. 
As we assumed the accretion disk to be geometrically thin and optically thick, we only considered the emission from the upper half of the NS surface. 
We considered the direct emission from the SL and neglected the components corresponding to the reflection of the SL emission from the accretion disk surface or scattering off the wind above the disk.

We first calculated the observed flux to estimate the polarization of the emission from the SL. 
For completeness, we repeated (with some adjustments for the SL geometry) the method that was used to compute the observed flux \citep[see also][]{PB06,SNP18,Bogdanov19L26,Suleimanov20} and then extended it to the description of the polarization following \citet{poutanen20}. 

To obtain the emission from the whole SL, we started by calculating the emission from a small surface element with colatitude $\theta$ and azimuthal angle $\phi$. 
Then, we integrated over the whole visible area of the layer in the nonrotating frame. 

We chose the static coordinate system so that the unit vector in the direction of the observer was $\unit{k}=(\sin i, 0, \cos i)$, where $i$ is the inclination of the spin axis to the line of sight (see Fig.~\ref{fig:geom} for the geometry).  
The unit vector of the normal to the surface element (which is parallel to the radius vector of the emission point in the case of a spherical star) is 
$\unit{n}=(\sin\theta \cos\phi, \sin \theta \sin\phi,\cos\theta)$. 
Thus, the angle between the surface element radius vector and the line of sight is given by  
\be \label{eq:psi}
\cos\psi \equiv \unit{k}\cdot\unit{n}
= \cos i\ \cos\theta+\sin i\ \sin \theta\ \cos\phi . 
\ee
Because photon trajectories in the Schwarzschild metric are planar, the unit vector of the photon momentum close to the surface can be obtained as a linear combination: 
\be \label{eq:k0}
\unit{k}_0 = \frac{\sin \alpha\ \unit{k} + \sin{(\psi - \alpha)}\ \unit{n}}{\sin{\psi}}, 
\ee
where $\alpha$ is the angle between the surface normal and the photon momentum, 
\be
\cos\alpha=\unit{k}_0\cdot\unit{n} . 
\ee
For our calculations, we used the approximate analytical formula for the light bending $\alpha(\psi)$ as suggested by \citet{B02}, 
\be
\cos\alpha \approx 1 - (1-u)\,  (1-\cos\psi),
\ee
where $u=R_{\rm S}/R$ is the NS compactness, with $R_{\rm S}=2GM/c^2$ being the Schwarzschild radius. 

The material in the SL moves with a velocity $v$ in the direction defined by the unit vector $\unit{v}=(-\sin\phi,\cos\phi,0)$.
For most of the models, we assumed a solid-body-like rotation profile in which the equatorial velocity is a fraction $\beta_0$ of the Keplerian velocity, 
\be\label{eq:beta_eq}
\beta (\theta) = \beta_{\rm eq} \sin\theta = \beta_0  \frac{v_{\rm Kepl}}{c} \sin\theta =  \beta_0 \ \sqrt{\frac{u}{2 \ (1-u)}} \sin\theta .
\ee
The angle between the velocity vector and the photon direction is then 
\be \label{eq:cosxi}
\cos\xi = \unit{v} \cdot \unit{k}_0 = \frac{\sin\alpha}{\sin\psi}\  \unit{v} \cdot \unit{k}=  -  \frac{\sin\alpha}{\sin\psi} \sin i\ \sin\phi\  .
\ee

The observed flux at energy $E$ from the surface element is
\be \label{eq:dFE}
\rmd F_E= I_E  \rmd\Omega_{\rm obs} ,
\ee
where 
\be \label{eq:dOmegaobs}
\rmd\Omega_{\rm obs} =  {\cal D} \cos \alpha \frac{\rmd S}{D^2} 
\ee  
is the solid angle of the surface element as observed from the distance $D$, with the layer element area $dS=R^2 d\cos\theta\, d\phi$ and the lensing factor
\be
{\cal D} =\frac{1}{1-u}\frac{d \cos \alpha}{d \cos \psi} 
\ee
being equal to unity in the Beloborodov approximation.
The observed specific intensity can be related to the intensity measured in the corotating frame as 
\be\label{eq:IE_EE}
I_{E} = \left (\frac{E}{E'}\right )^3 I'_{E '} (\alpha',\theta) ,
\ee
where $\alpha'$ is the angle of the photon momentum to the surface normal as measured in the corotating frame of the surface element, and all the quantities in the corotating frame are labeled with a prime. 
The ratio of the energies combines gravitational redshift and the Doppler effect,
\be \label{eq:EEpr_sph}
 \frac{E}{E'} = \delta \sqrt{1-u} .
\ee
The Doppler factor 
\be \label{eq:Doppler_sph}
\delta = \frac{1}{\gamma(1-\beta\cos\xi)} 
\ee
depends on the SL velocity at this colatitude relative to the external nonrotating frame, and the Lorentz factor is 
\be  \label{eq:Lorentzgamma}
\gamma(\theta)=\frac{1}{\sqrt{1-\beta^2(\theta)}}. 
\ee
The angle $\alpha'$ is related to the analogous angle measured in the static frame as \citep{PG03} 
\be
\cos\alpha'=\delta\cos\alpha .
\ee 
Assuming that radiation from the SL can be described by electron-scattering dominated semi-infinite atmosphere \citep{Cha60, Sob63}, we approximated the energy and angular distribution of the specific intensity as  \citep{Suleimanov20}
\be\label{eq:angular_fraction}
I'_{E'}(\mu,\theta) = \frac{1}{f^4_\mathrm{c}} B_{E'}(f_\mathrm{c} T_{\mathrm{eff}}) a_{\rm es}(\mu) ,
\ee 
where $a_{\rm es}(\mu)\approx 0.421 + 0.868 \mu$, $\mu=\cos\alpha'$, $B_E$ is the Planck function, 
$T_{\mathrm{eff}}$ is the effective temperature, which can be a function of colatitude $\theta$, and the color correction $f_\mathrm{c}$ is assumed to be latitude-independent. 
We finally obtained the observed flux from the surface element
\be
\rmd  F_{E} = \frac{R^2}{D^2} \cos \alpha\ {\cal D} \ \left(\delta\sqrt{1-u}\right)^3  I'_{E'} (\alpha',\theta) \,\rmd\cos\theta\, \rmd\phi. 
\ee

To describe the polarimetric properties of the emission from a surface element, we introduced the Stokes vector,
\be 
\rmd  \bm{F}(E)= \left(\begin{array}{c}  \rmd F_I(E) \\ \rmd F_Q(E) \\ \rmd  F_U (E) \end{array}\right)  = \rmd  F_E \left(\begin{array}{c}  1  \\ P\cos2\chi \\ P \sin2\chi  \end{array}
\right) ,
\ee 
where $P$ is the observed PD, and $\chi$ is the observed PA, both of which depend on energy. The fourth component of the Stokes vector describing circular polarization was removed from consideration, as circular polarization is conserved along the photon trajectory in the absence of the plasma effects and is not observable with \ixpe or \extp.
Because the linear PD is also conserved along the photon trajectory, we set $P$ to be equal to the  PD of the emitted radiation at the NS surface, corresponding to the optically thick electron-scattering-dominated plane-parallel atmosphere \citep{Cha60,VP04}, 
\be \label{eq:ChandrasekharPD}
P_{\mathrm{es}} (\mu) =  -  \frac{1-\mu}{1+3.582 \mu} \  0.1171. 
\ee
The corresponding PA $\chi$ was calculated as in \cite{poutanen20},
\be \label{eq:tanchi_rel_bend}
\tan \chi = 
\frac{ \sin \theta  \sin \phi + \beta \ A  } 
{-\sin i \cos \theta + \cos i \sin \theta \cos\phi  -  \beta\sin\phi \ C }, 
\ee    
where 
\beq  \label{eq:aux_rel_bend}
A&=& \frac{\sin\psi}{\sin\alpha} B+  \frac{\cos\alpha-\cos\psi}{\sin\alpha \sin\psi} (\cos\phi-B\cos\psi) , \nonumber \\ 
B&=& \sin i \sin \theta+\cos i \cos \theta \cos\phi, \\
C&=& \frac{\sin\psi}{\sin\alpha}\cos \theta +  \frac{\cos\alpha-\cos\psi}{\sin\alpha \sin\psi} (\cos i -\cos \theta \cos\psi) .  \nonumber
\eeq

Finally, to obtain the polarized emission from the whole SL, we integrated over its visible area as defined by the visibility condition $\cos\alpha>0$ \citep{Suleimanov20}, 
\begin{eqnarray} 
\label{eq:FStokes}
 \bm{F}(E) &=& \left(\begin{array}{c}  F_I(E)  \\ F_Q(E) \\ F_U (E) \end{array}\right)  =   \frac{R^2}{D^2}  
\iint\displaylimits_{\cos \alpha>0}
\cos \alpha\ {\cal D} \ \left(\delta\sqrt{1-u}\right)^3 \nonumber \\  
&\times& I'_{E '} (\cos\alpha',\theta) 
\left(\begin{array}{c}  1  \\ P_{\rm es}(\cos\alpha')\cos2\chi \\ P_{\rm es}(\cos\alpha') \sin2\chi  \end{array}
\right)
\rmd\cos\theta\ \rmd\phi.   
\end{eqnarray}  
Both the PD ($P_{\rm es}$) and PA ($\chi$) in this equation depend on the coordinates of the surface element $(\theta, \phi)$.
For the whole SL emission, the observed PD and PA at every energy were computed via 
\be \label{eq:Pobs}
P_{\mathrm{obs}} = \frac{\sqrt{F_{Q}^2+F_{U}^2}}{F_{I}}  
\ee
and
\be \label{eq:chiobs}
\cos(2\chi_{\mathrm{obs}}) = F_{Q}/(P_{\mathrm{obs}}F_{I}) , \quad 
\sin(2\chi_{\mathrm{obs}}) = F_{U}/(P_{\mathrm{obs}}F_{I}) .  
\ee

\section{Results for various geometries}
\label{sec:Results}

\begin{figure*}
\begin{center}
\includegraphics[width=1.\textwidth]{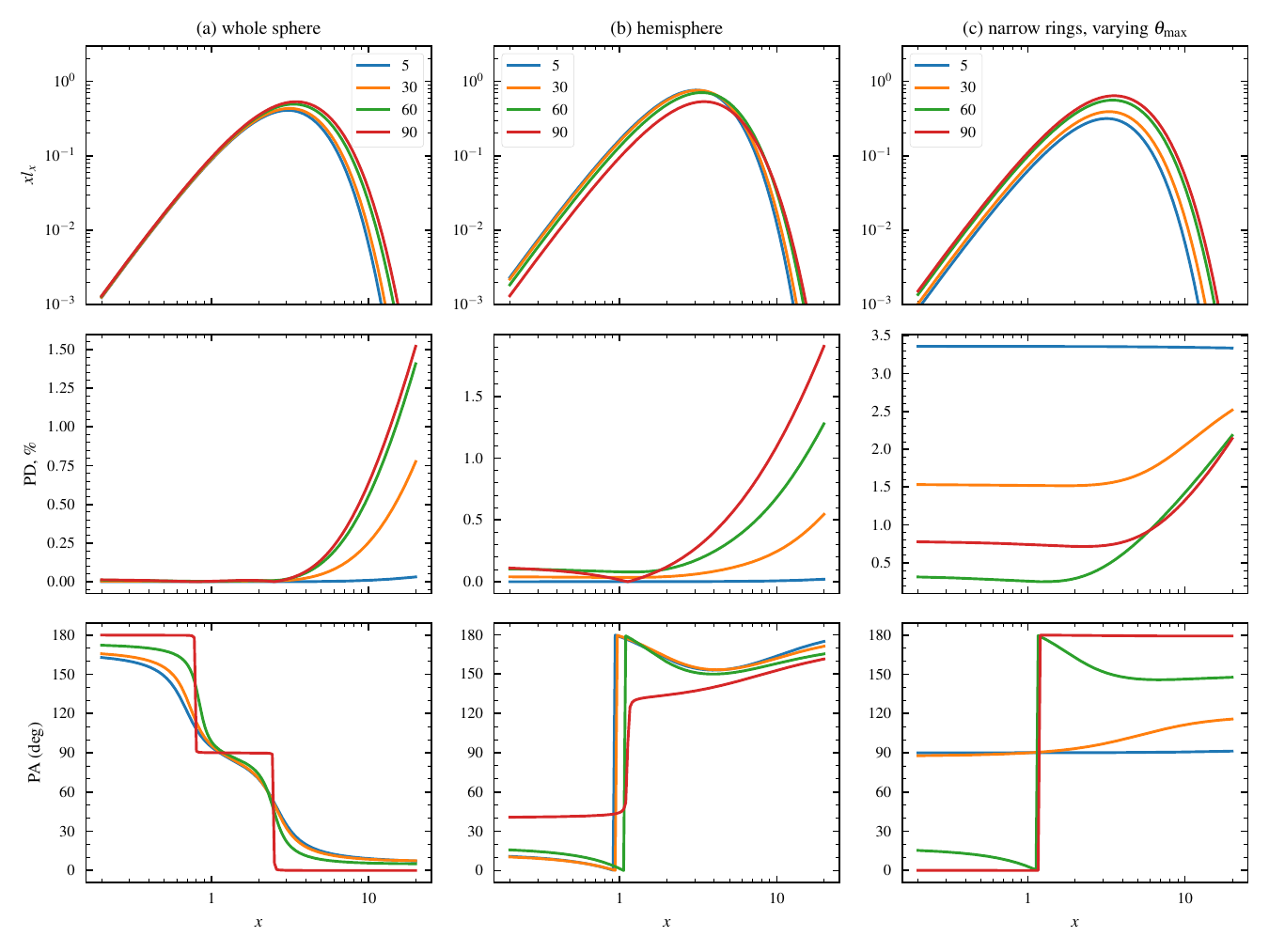}
\caption{Dimensionless luminosity $xl_{x}$, PA, and PD as functions of the dimensionless photon energy, $x = E/f_{\rm c}k_{\rm B}T_{\rm eff}$. For all the cases presented, $\beta=0.35$.   
Results for the whole sphere (column a) and upper hemisphere (column b), observed at inclinations of 5\degr, 30\degr, 60\degr, and 90\degr\ are shown with blue, orange, green, and red lines, respectively. 
Results for 1\degr-wide rings  observed at an inclination 90\degr\ are shown in column (c) with blue, orange, green, and red lines for ring colatitude of 5\degr, 30\degr, 60\degr, and 90\degr, respectively.}
\label{fig:Fig2.5}
\end{center}
\end{figure*}

\begin{figure*}
\begin{center}
\includegraphics[width=1.0\textwidth]{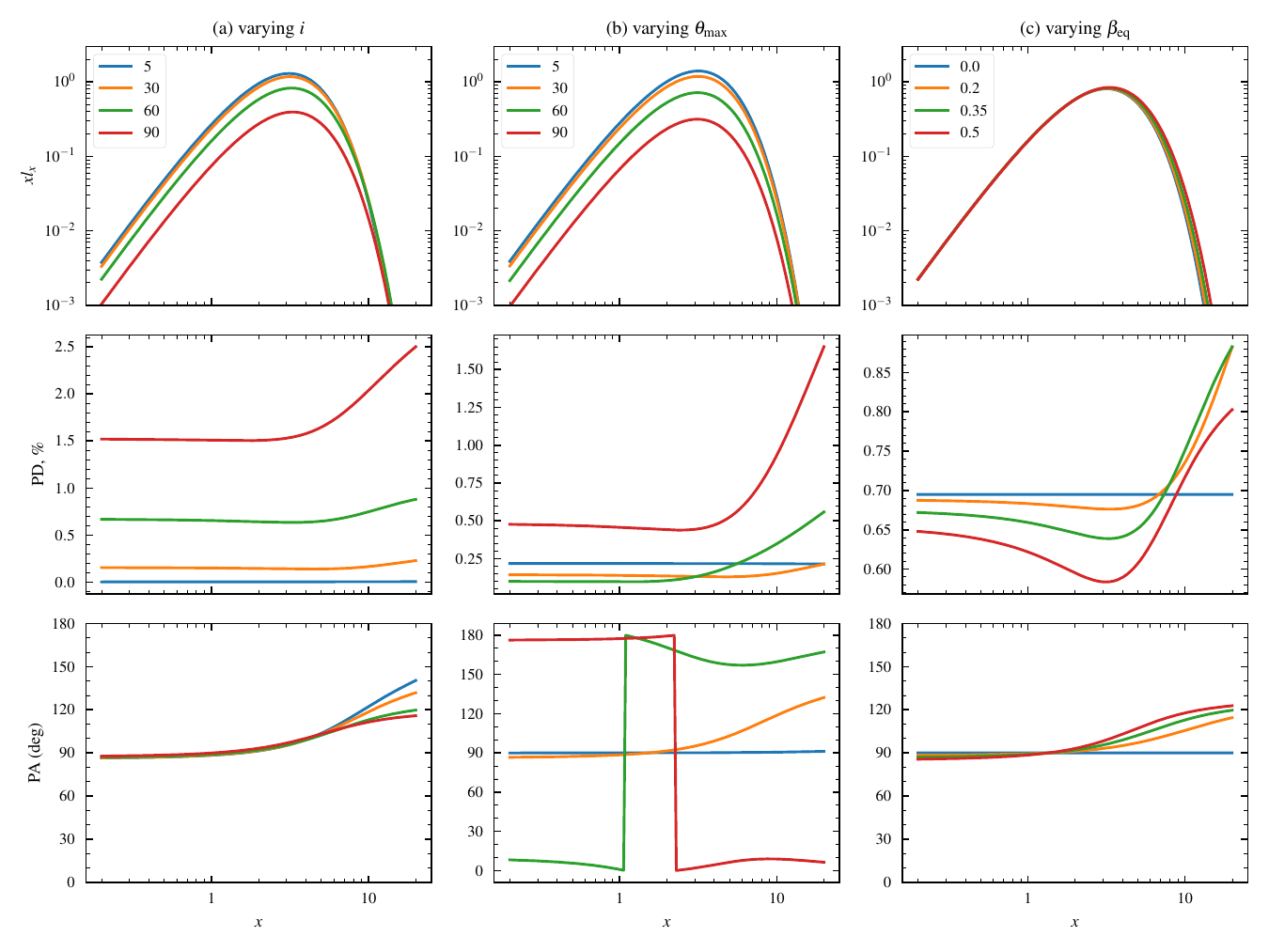}
\caption{Dimensionless luminosity $xl_{x}$, PA, and PD as functions of the dimensionless photon energy, $x = E/f_{\rm c}k_{\rm B}T_{\rm eff}$ for emission from the 1\degr-wide rings with the fiducial parameter set $\theta_{\rm max} = 30\degr$, $i = 60\degr$, and $\beta_{\rm eq} = 0.35$. Column (a) shows the results for various inclinations with other parameters from the fiducial set. The blue, orange, green, and red lines correspond to $i=$5\degr, 30\degr, 60\degr, and 90\degr, respectively. Column (b) shows the results for various ring colatitudes with other parameters from the fiducial set. The blue, orange, green, and red lines correspond to $\theta_{\rm max}$=5\degr, 30\degr, 60\degr, and 90\degr, respectively. Column (c) shows the results for various equatorial velocities with other parameters from the fiducial set. The blue, orange, green, and red lines correspond to $\beta_{\rm eq}$=0.0, 0.2, 0.35, and 0.5, respectively. The PA is zero for vertical polarization and is measured counterclockwise.} 
\label{fig:Fig3}
\end{center}
\end{figure*}

\begin{figure*}
\begin{center}
\includegraphics[width=1.0\textwidth]{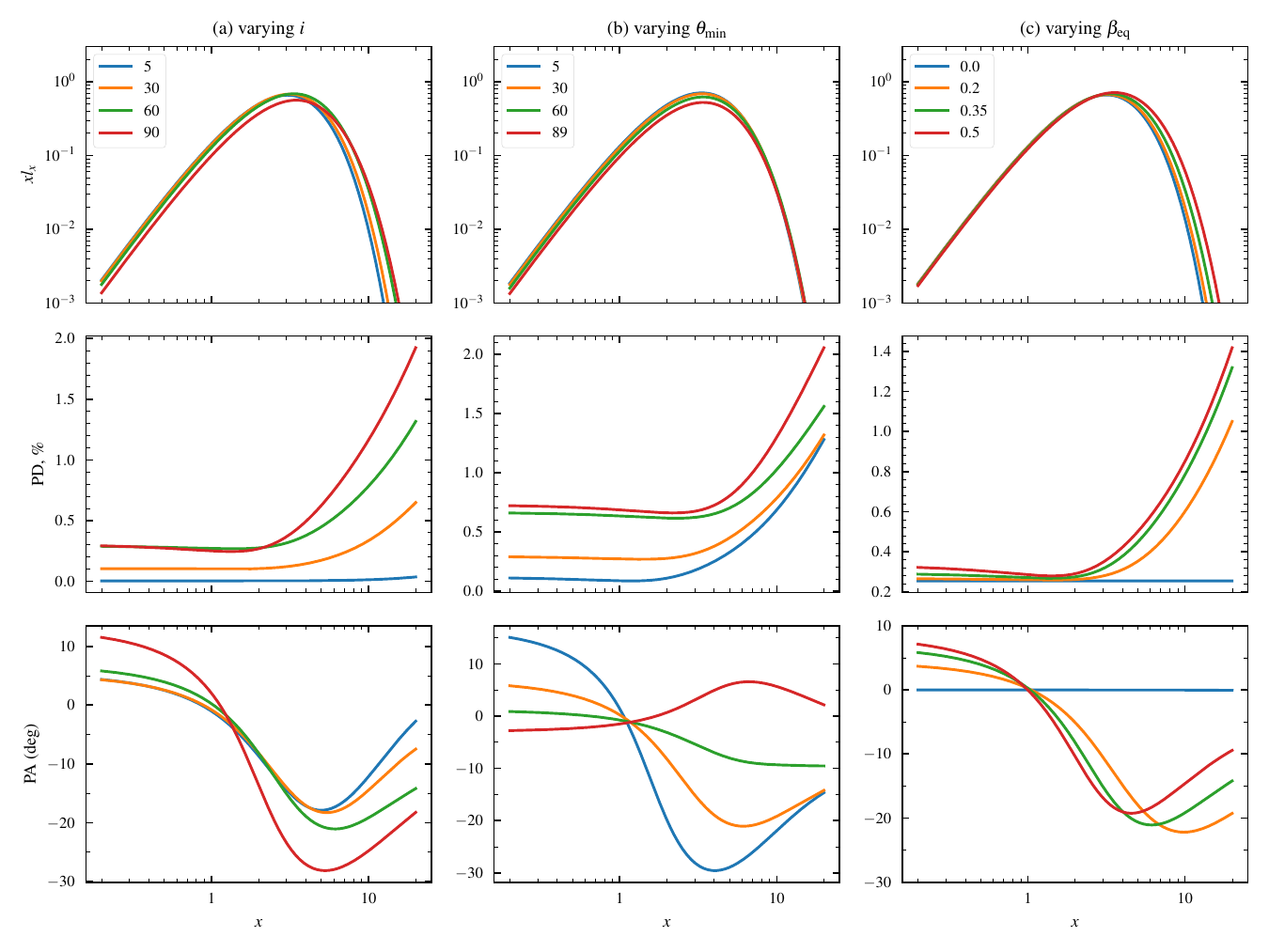}
\caption{Same as Fig.~\ref{fig:Fig3}, but for emission from the broad SL. The fiducial set of parameters is $\theta_{\rm min} = 30\degr$, $\theta_{\rm max} = 90\degr$, $i = 60\degr$, and $\beta_{\rm eq} = 0.35$.}
\label{fig:Fig4}
\end{center}
\end{figure*}

The formalism derived in Sect.~\ref{sec:theory} was  applied to the various geometries of the SL. 
The resulting polarimetric characteristics were expressed as a function of the dimensionless photon energy 
\be
x = \frac{E}{f_{\rm c} k_{\rm B} T_{*}},
\ee
where $k_{\rm B}$ is the Boltzmann constant and $T_{*}$ is a characteristic effective temperature $T_{\rm eff}$ of the SL. 
By scaling Eq.~\eqref{eq:FStokes} for the observed flux $EF_E$ by the typical flux at the NS surface $\sigma_{\rm SB}T_{*}^4$, we introduced the dimensionless luminosity (Stokes vector) as
\begin{eqnarray}
\label{eq:lum}
 x\, \bm{l}_x\!\! &=&\!\!  \frac{D^2}{\Delta R^2}   \frac{E\bm{F}_E}{\sigma_{\rm SB}T_{*}^4} = \frac{15}{\pi^5} \!\!\iint_S\!\! \cos \alpha\ {\cal D} \ \left(\delta\sqrt{1-u}\right)^4 a_{\rm es}(\cos\alpha') \nonumber   \\
    &\times&    \frac{x'^4}{{\rm e}^{x'T_{*}/T_{\rm eff}(\theta)}-1}
\left(\begin{array}{c}  1  \\ P_{\rm es}(\cos\alpha')\cos2\chi \\ P_{\rm es}(\cos\alpha') \sin2\chi  \end{array}
\right)
\rmd\cos\theta\ \rmd\phi. 
\end{eqnarray}
Here, $\Delta$ is the fraction of the NS surface occupied by the SL, and $x'=x/(\delta \sqrt{1-u})$.
The dependences of the dimensionless luminosity $x l_x$, PD, and PA on dimensionless energy $x$ for various SL models are shown in Figs.~\ref{fig:Fig2.5}--\ref{fig:Fig5}.  
We note that for an isothermal ($T_{\rm eff}(\theta)=T_{*}$) nonrotating NS surface, Eq.~\eqref{eq:lum} is reduced to 
\begin{equation}
x\, \bm{l}_x = \frac{15}{\pi^4} (1-u)^{-1} \frac{x^4}{{\rm e}^{x/\sqrt{1-u}}-1} \left(\begin{array}{c}  1  \\ 0 \\ 0  \end{array}
\right) .
 \end{equation}

\subsection{Basic geometries: Sphere, hemisphere, and one-degree rings}

We started with the most basic geometrical models and explored the emission from a SL that is shaped as a sphere, hemisphere, and a narrow ring of 1\degr\ width. 
We also assumed that the effective temperature is constant over the SL.
Studying these basic cases can support our general understanding of the SL emission and its polarization. The equatorial velocity of the plasma flow in the SL $\beta_{\rm eq}$ was fixed at the value of 0.35, which corresponds to $\beta_0=0.68$ for a NS with a radius $R=12$~km and a mass $M=1.4~M_\odot$. 

We first studied the emission from the whole sphere (Fig.~\ref{fig:Fig2.5}a) for different inclination angles noted in the plot. Except for the energies above the peak in the luminosity, the emission is almost not polarized, regardless of the inclination, because the problem is nearly symmetric. The increase in PD is connected to the exponential cutoff in the spectrum of the SL and the Doppler effect. The rotation of the SL increases the observed energy of the light from the part of the SL in which the matter flows toward the observer. As the light from that part of the SL is polarized, the PD increases. This pattern is also more visible for higher inclinations because the direction of the matter flow is parallel to the NS equator. The variation in the PA with energy is most peculiar: in the extreme case of an inclination $i=90\degr$ (red lines), three steps appear, and for the lower inclinations, the pattern is smoothed. These steps also appear due to the Doppler effect: at lower energies, the emission from the part of the sphere that moves away from the observer dominates, and the light from this area is polarized perpendicular to the disk plane, so that PA=$0\degr$ (or $180\degr$). For medium energies, the flux is dominated by the observed middle section of the NS from the poles to the equator. This light is polarized in the disk plane, so that PA=$90\degr$. At higher energies, the emission from the part of the sphere where matter flows toward the observer is observed. The polarization is again perpendicular to the disk plane, and the PA is zero. 

A hemisphere is a good approximation for the presence of the accretion disk because it covers half of the NS (Fig.~\ref{fig:Fig2.5}b). In this case, PD is nonzero even at the lowest photon energies. The most significant change between the two cases is in the PA. We again examine the edge-on case (red lines). In general, the polarization from the electron-scattering-dominated optically thick cold slab is parallel to the surface of the slab. For the whole sphere and counterclockwise rotation of the matter, the left half of the sphere is brighter because of the Doppler effect. As this case is otherwise symmetric, the overall polarization is expected to be dominated by the PD of the brighter part and perpendicular to the disk plane for any energies, so that PA is zero (or 180\degr). For the hemisphere case, however, the quarter of the sphere now dominates the overall picture. 
At lower energies, the emission is therefore dominated by the right part of the star, so that the PA shifts toward 45\degr. Moreover, PA rotates rapidly close to the energy $x\approx1$ (i.e., $E\approx kT_{\rm eff}f_{\rm c}$).
At higher energies, the PA rotates toward $\approx140\degr$ (which is equivalent to $\approx-40\degr$), and PD rises because the emission is now is dominated by Doppler boosting of a smaller portion on the left part of the star, where the matter moves toward to the observer.

For the last column of the figure, we considered a narrow ring with a width of $1\degr$ at different colatitudes and report the value of the colatitude of the lower edge of the ring, $\theta_{\rm max}$ (Fig.~\ref{fig:Fig2.5}c).
With the inclination fixed at $90\degr$, the PD decreases with the colatitude of the ring (except for the case of the ring at $\theta_{\rm max}=60\degr$, which is discussed below). 
This occurs because the emission observed from the normal to the surface is not polarized. For instance, the observer at an inclination of $90\degr$ sees the radiation from the polar ring polarized (practically) horizontally. At the equator, on the other hand, the emission that arrives at the observer from the part of the ring in which the azimuthal angle is close to the angle of the observer is not polarized at all, and the light from this part of the SL contributes to a further decrease in the overall observed PD. 
This trend of a decrease in the PD with colatitude of the ring, however, breaks for the higher colatitudes because the PD from the ring at the equator exceeds the PD from the ring at a colatitude of $60\degr$. This is because the PA of the light from the ring observed at this intermediate angle varies with the azimuthal angle (i.e., the light is not horizontally polarized, like at the pole, nor vertically, like at the equator), leading to partial cancellation of the overall PD. 
This is also reflected in the PA-dependence plot. 

\subsection{Narrow one-degree rings: Detailed study}

We explored the emission from rings of 1\degr\ width in detail. Unless specified otherwise, our fiducial set of parameters was $\theta_{\rm max} = 30\degr$, $i = 60\degr$, and $\beta_{\rm eq} = 0.35$.
As we understood the results of the simple cases shown in Fig.~\ref{fig:Fig2.5}, with the more complicated cases, we relied on the results of the simulations and switched to the description and interpretation rather than an explanation of the phenomena presented in the plots. 

Figure~\ref{fig:Fig3}a illustrates the differences in the spectra for different inclinations. As in the case considered in Fig.~\ref{fig:Fig2.5}c, an increase in the angle between the normal to the surface and the direction to the observer increases the PD. Additionally, unpolarized emission is observed when the observer is close to the pole because the entire ring is seen and the problem becomes (almost) azimuthally symmetric around the direction to the observer. The change in the inclination does not affect the PA in this case. The differences only occur at higher energies because of the Doppler effect. 

The effect of varying the colatitude of the ring is shown in Fig.~\ref{fig:Fig3}b. The Doppler effect again affects the spectrum of the ring more prominently when the ring is closer to the equator, where the gas moves with the highest velocities. We also note that the lowest PD is expected from the ring at the colatitude $\theta_{\rm max} = 60\degr$, which coincides with the inclination in this scenario, as PD turns to zero when the emitting region is observed along the  normal. 

For the cases presented in Fig.~\ref{fig:Fig3}c, we changed the equatorial velocity. In the nonrotating case (blue line), PD and PA do not depend on energy. A slight decrease in the PD with increasing velocity is observed at lower energies, while an increase is seen at higher energies. 
In all three considered cases of a rotating SL, the rotation of the PA reached 30\degr.

\begin{figure*}
\begin{center}
\includegraphics[width=0.70\textwidth]{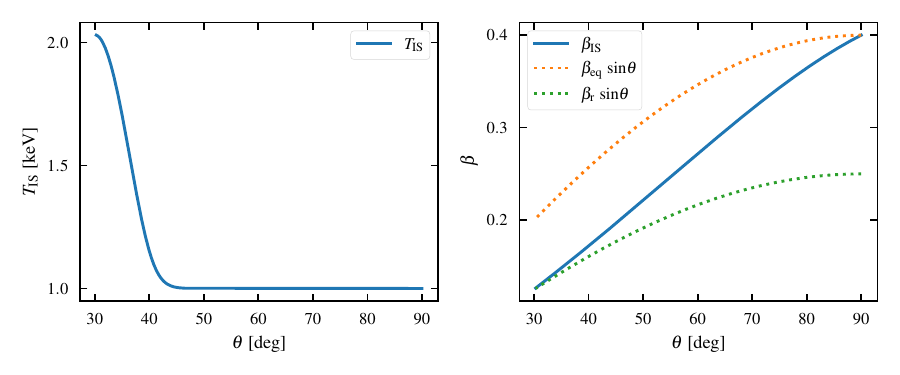}
\caption{Effective temperature (left panel) and velocity of the matter flow (right panel) for the Inogamov-Sunyaev SL as functions of the colatitude. The parameters are $\theta_{\rm peak} = 30\degr$, $\beta_{\rm r}=0.25$,  and $\beta_{\rm eq}=0.4$. }
\label{fig:IS_parameters}
\end{center}
\end{figure*}

\subsection{Wide spreading layer}

For the next step of our simulations, we broadened the 1\degr\ ring into a wider layer. Unless specified otherwise, we consider in this section an isothermal SL with $\theta_{\rm min} = 30\degr$, $\theta_{\rm max} = 90\degr$, $i = 60\degr$, and $\beta_{\rm eq} = 0.35$. Similar to Fig.~\ref{fig:Fig3}, we took for Fig.~\ref{fig:Fig4}a the fiducial set and examined the influence of the inclination on the observed emission for different inclinations between 5\degr\ and 90\degr. 
We again observe that the PD is the highest for $i=90\degr$. Observing the system virtually from the pole ($i=5\degr$) shows the lowest PD, as the SL appears nearly axially symmetric around the direction to the observer. 
For the case shown in Fig.~\ref{fig:Fig4}b, we investigated the dependence on the width of the SL by changing the $\theta_{\rm min}$ from 5\degr\ to the values mentioned in the plot. Our dimensionless luminosity was normalized by the surface area of the emitting region, so that the observed flux did not change much with the width of the layer. However, the PD decreased when the SL changed from a narrow belt to a hemisphere. Finally, we changed in Fig.~\ref{fig:Fig4}c the value of the equatorial velocity of the plasma in the SL. The change in this parameter predominantly affected the flux at the highest energies. The difference in the PD and PA between a nonrotating case (blue line) and rotating at the Keplerian velocity case (red line) is significant. In all three cases presented in Fig.~\ref{fig:Fig4}, we obtained a change of up to $40\degr$ in the PA.

\section{Results for various physical configurations}
\label{sec:Results_em}

In the previous section, we considered various geometrical configurations of both SL and the binary system. To do this, we kept the surface temperature, the velocity of the relativistic flow, and the polarization calculations as simple as possible. However, a homogeneous effective surface temperature across the whole accretion flow is not realistic, and neither is the weak dependence of the velocity on $\theta$ presented in Sect.~\ref{sec:theory}. Chandrasekhar's limit on the polarization from an infinitely thick slab can also be replaced with exact Thomson scattering calculations. We searched for alternative models and tested the polarimetric properties of these SLs.

\subsection{Inogamov-Sunyaev spreading layer} 

\citet{IS99} introduced the concept of a SL that is most luminous at the edge of the layer, where it reaches the surface of the NS, decelerates, and emits, rather than at the stellar equator, where the gas levitates because its velocity is close to the Keplerian velocity. To match the suggested flux profile \citep[as shown in Fig.~8 in][]{IS99}, we introduced the dependence of the effective temperature on colatitude,
\be\label{eq:T_IS}
T_{\rm IS}^4 (\theta) = H(\theta-\theta_{\rm peak}) \ \left\{T_{\rm min}^4 + T_{*}^4 \exp{\left[- \frac{(\theta-\theta_{\rm peak})^2}{2\sigma^2}\right]}\right\} ,
\ee
where $H$ is the Heaviside step function, $T_{\rm min}$ is the temperature of the SL off-peak, $T_{*}$ is a characteristic temperature impacting the height of the peak, and $\sigma$ parameterizes the width of the peak. This dependence is illustrated in the left panel in Fig.~\ref{fig:IS_parameters}. We assumed a Gaussian peak at $\theta_{\rm peak} = 30\degr$ with a full width at half maximum of 20\degr\ ($=2.355 \sigma$), $T_{\rm min}=1$~keV, and $T_{*}=2$~keV.

\begin{figure*}
\begin{center}
\includegraphics[width=1.0\textwidth]{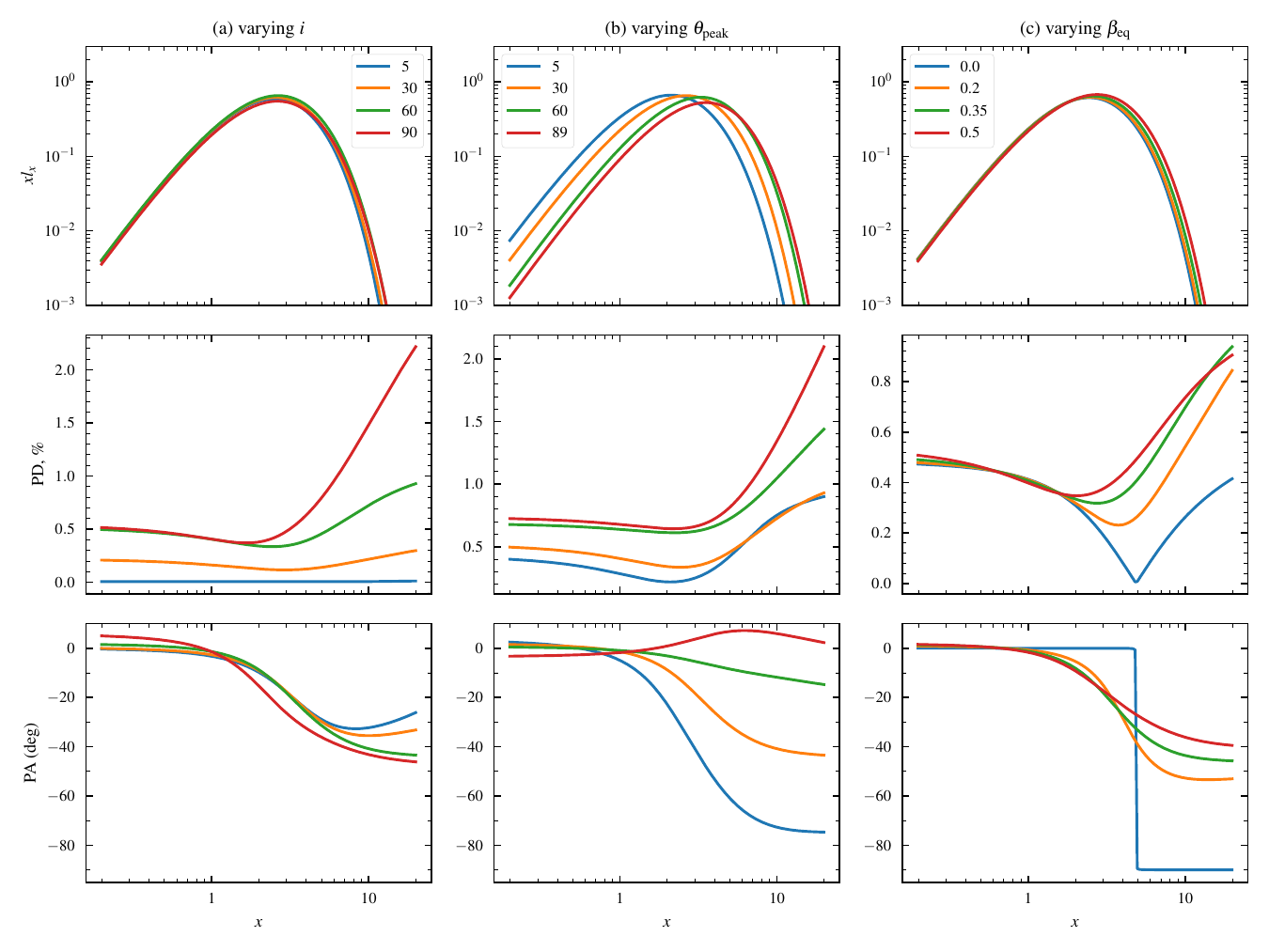}
\caption{Spectropolarimetric characteristic of the Inogamov-Sunyaev SL model. 
The upper panels show the dimensionless luminosity $xl_{x}$, and the middle and lower panels show the PD and PA as functions of the dimensionless photon energy  $x = E/f_{\rm c}k_{\rm B}T_{*}$. 
The fiducial parameter set is $\theta_{\rm peak} = 30\degr$, $i = 60\degr$, $\beta_{\rm r} = 0.25$, and $\beta_{\rm eq} = 0.4$. 
Column (a) shows the results for various inclinations. The blue, orange, green, and red lines correspond to $i=$5\degr, 30\degr, 60\degr, and 90\degr, respectively. Column (b) shows the results for various colatitudes of the peak $\theta_{\rm peak}$. The blue, orange, green, and red lines correspond to $\theta_{\rm peak}=5\degr$, 30\degr, 60\degr, and 89\degr, respectively. Column (c) shows the results for various equatorial velocities. The blue, orange, green, and red lines correspond to $\beta_{\rm eq}=$0, 0.2, 0.35, and 0.5, respectively. In this case, the edge velocity was fixed at $\beta_{\rm r} = 0.5 \beta_{\rm eq}$.}
\label{fig:Fig5}
\end{center}
\end{figure*}

As the gas at the equator was assumed to levitate with a velocity corresponding to that of the accretion disk and to slow down closer to the edge of the SL, we also introduced a stronger dependence of the velocity on colatitude, 
\be\label{eq:beta_IS}
\beta_{\rm IS} (\theta) = \left(\beta_{\rm eq}- \frac{\cos\theta}{\cos\theta_{\rm peak}} \ (\beta_{\rm eq}-\beta_{\rm r})\right) \sin\theta ,
\ee
where $\beta_{\rm eq}$ is the velocity of the SL surface at the equator of the NS, and $\beta_{\rm r} \sin{\theta_{\rm peak}}$ is the velocity of the SL surface at the edge of the SL that is associated with the rotation of the NS. This dependence is illustrated in the right panel in Fig.~\ref{fig:IS_parameters}. When these dependences are taken into account, the dimensionless luminosity given in Eq.\,(\ref{eq:lum}) changes accordingly.

The results are presented in Fig.~\ref{fig:Fig5}. We varied the parameters of the SL in the same way as for Figs.~\ref{fig:Fig3} and \ref{fig:Fig4}. Overall, the behavior is quite similar to the one shown in Fig.~\ref{fig:Fig4}. The PD follows the same trends  with the energy and varied parameters. The PA is $\approx0\degr$ at lower energies and decreasing with energy. We note a stronger dependence of the PA on energy in this case. The most significant difference from Fig.~\ref{fig:Fig4} is in Column (c), where we study the impact of varying the SL velocity on the polarimetric properties. In the case of $\beta_{\rm eq}=0$, the vertical polarization at lower energies changes to horizontal polarization at higher energies. This occurs because the temperature of the layer is a function of colatitude. The emission from the colatitudes above $\approx45\degr$ has an effective temperature of 1 keV, while the emission from lower colatitudes is hotter, with $T_{\rm eff}$ exceeding 2 keV at the SL edge. The light from these parts of the SL is observed in different energy ranges: emission from $\theta>45\degr$ is seen at lower energies, and the emission of the SL edge is in the higher-energy range. As we already noted in Sect.~\ref{sec:Results} and especially Column (b) of Fig.~\ref{fig:Fig3}, the polarization from the rings close to the equator is vertical, while the polarization from the lower colatitudes, for instance, $\theta\approx30\degr$, is horizontal. The increased velocity of the SL causes the energy dependence of the PD to become smoother, but the decrease in PD associated with the fast rotation of the PA is still visible.

\subsection{Optically thin spreading layer}

\begin{figure} 
\begin{center}
\includegraphics[width=0.9\linewidth]{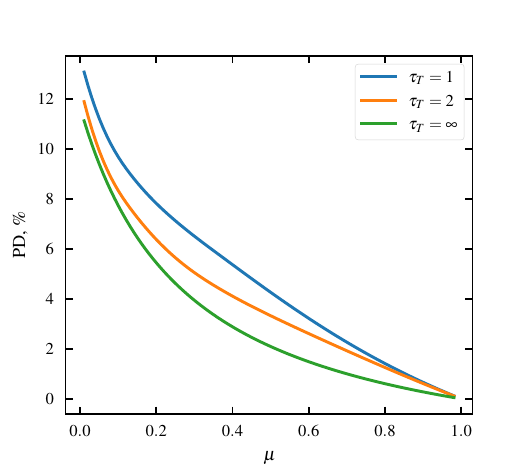}
\caption{PD as a function of emission angle $\mu = \cos \alpha'$ for various SL optical thicknesses. 
}\label{fig:thinPDs}
\end{center}
\end{figure}

In the previous examples, we assumed the SL to be optically thick and used Eq.\eqref{eq:ChandrasekharPD} for the PD as a function of emission angle. We can instead assume that the SL is optically thin and perform the exact calculation of the Thomson scattering in the SL to estimate the polarization of the emission. The formalism for calculating the polarization from the Thomson scattering was discussed in \citet{1985A&A...143..374S} and \citet{VP04}. From our calculations, we learned that the PD as a function of emission angle barely depends on the optical thickness of the layer. Emission is still polarized in the plane of the SL surface, and the effects of the first few scatterings cancel out, leading to depolarization, as shown in Fig.~\ref{fig:thinPDs}.

High polarization from Compton scattering is expected if we were only to observe the photons after they have experienced many scatterings. This is the case in hot Comptonizing regions, where cool seed photons are upscattered to the X-ray range, and we only see multiple scattered photons in the \ixpe range 2--8 keV. This is not the case for the WMNSs, as we expect the SL to be at a low temperature of 2--3 keV. In this case, we observe all the photons in the same energy range, regardless of the number of scatterings they experience. As is shown, for instance, in Fig.~2 of  \cite{VP04}, the PD of the photon changes sign (i.e. PA rotates by 90\degr) depending on the number of scatterings it experienced. When we see them all in the same energy range, we therefore expect an overall low polarization regardless of the optical depth. 

\section{Applications} \label{sec:IXPE}

In the past two years, \ixpe has observed a dozen WMNSs. The model presented in this paper gave valuable insights that supported the interpretation of the observation results. First, we learned from the simulations that the emission of the SL is polarized perpendicular to the accretion disk, while for the optically thick electron-scattering-dominated disk,  the polarization vector lies in the disk plane. In the spectrum of the WMNSs, we usually see two main components, the softer of which is interpreted as the emission from the disk, while the harder component comes from the BL or the SL. Thus, if a source presents a strong dependence of the PA on energy, we assumed that the harder component of the spectrum comes from the SL rather than the BL, which produces polarization similar to that of the disk. This was the case, for instance, in the \ixpe  data on \mbox{Cyg~X-2} \citep{Farinelli23}. However, in this source, the estimated PD from the harder component was about 4\%, which is much higher than the PD in our simulations. Similarly, in \mbox{4U~1820$-$303} \citep{DiMarco23}, the shift by 90\degr\ in the PA suggests the presence of a SL, but PD at the higher energies reaches 10\%, which is far above our predictions. 

On the other hand, in the cases of \mbox{Sco~X-1} \citep{LaMonaca2024} or \mbox{GX~13+1} \citep{Bobrikova24, Bobrikova24b}, a clear absence of PA rotation with energy suggested that the SL, at least in our model, cannot produce the harder component of the spectrum. The most peculiar case of the \mbox{Cir~X-1} \citep{Rankin2024} required a complicated explanation of the observed phenomena, but the authors concluded that the Comptonized component of the spectrum might come from the SL in the hard state of the source, while in the soft state, the Comptonized emission might come from the BL. In sources \mbox{GX~9+9} \citep{Ursini2023}, \mbox{GX~5$-$1} \citep{Fabiani24}, and \mbox{XTE~J1701$-$462} \citep{Cocchi23}, a misalignment in the PA of the two components was found, but the statistics did not allow us to properly constrain the shift in the PA. There is no clear conclusion about the geometry of these sources. 

The model supports the data interpretation, but it is not fully capable of explaining the observed phenomena, even when an assumption about the presence of the SL in the source is made. 
Thus, additional sources of polarized emission are needed. 
For example, radiation produced close to the NS surface can be further scattered in the wind above the disk  \citep{Tomaru2024,Nitindala2025} or be reflected from the disk \citep{Lapidus1985}, which could lead to a higher PD of the emission. Observational results leave much space for further improvements to the current SL model.

\section{Summary}
\label{sec:Summary}

We developed a theoretical model for the emission from the SL of the WMNS. 
We derived exact analytic expressions for the Stokes parameters of the emission. We accounted for the special relativistic effects and light bending when calculating the PA. 

We computed the emission from the source with various geometries of the SL, and we accounted for the velocity of matter. We studied the impact of the inclination, matter velocity, and the geometrical configuration of the SL on the polarization of the observed emission. We calculated the polarization from the Inogamov-Sunyaev SL, where both the velocity of the matter and the intensity of the emission are functions of colatitude. 

We showed that for all the tested scenarios, the PD of the emission does not exceed 1.5\%. Recent observations of the \ixpe satellite showed that WMNSs emit light with stronger polarization, especially at higher energies. Hence, the SL, at least the current model, cannot be the only mechanism to explain the data. A higher polarization of the harder component of the WMNS spectrum can come from the reflection of the SL emission from the disk, scattering of the SL emission in the wind above the accretion disk, the emission from the jet, and others. Developing the emission models for these processes is the next step for further investigation of WMNSs. We also note the significant variability of the PA with energy that is observed in the emission of the SL. 

The developed model can support the \ixpe data analysis, and it was successfully used to interpret the already existing results of \ixpe. Combined with other tools and methods, it can shed new light on the geometry of the WMNSs and their emission mechanisms. 

\begin{acknowledgements}
This research was supported by the Academy of Finland grant 333112, and the grants 002200175 and 00240328 of the Finnish Cultural Foundation (AB).  
We thank the referee for their useful suggestions. 
\end{acknowledgements}

\bibliographystyle{aa}
\bibliography{ns}

\end{document}